\documentclass{article}
\usepackage{titlesec}
\titleformat{\paragraph}
{\normalfont\normalsize\bfseries}{\theparagraph}{1em}{}
\titlespacing*{\paragraph}
{0pt}{3.25ex plus 1ex minus .2ex}{1.5ex plus .2ex}
\usepackage{amsfonts}
\usepackage[english]{babel}
\usepackage{amsmath,txfonts}
\usepackage{float}
\usepackage{graphicx}
\usepackage{url}
\usepackage[colorlinks=true, allcolors=blue]{hyperref}

\newtheorem{theorem}{Theorem}

\newtheorem{example}[theorem]{Example}

\begin{document}

\title{ChatGPT in Linear Algebra: \\Strides Forward, Steps to Go}

\maketitle

\begin{tabular}{c c c}
Eli Bagno & Thierry Dana-Picard & Shulamit Reches \\
bagnoe$@$gmail.com & ndp$@$jct.ac.il & shulamit.reches$@$gmail.com\\

\\
\quad & Jerusalem College & \quad \\
\quad & of Technology & \quad \\
\quad & Jerusalem, Israel &\quad
\end{tabular}

\begin{abstract}
As soon as a new technology emerges, the education community explores its affordances and the possibilities to apply it in education. In this paper, we analyze sessions with ChatGPT around topics in basic Linear Algebra. 
We reflect the process undertaken by the ChatGPT along the recent year in our area of interest, emphasising the vast improvement that has been done in grappling with Linear Algebra problems. 
In particular, the question whether this software can be a teaching assistant or even somehow replace the human teacher, is addressed. As of the time this paper is written, the answer is generally negative. For the small part where the answer can be positive, some reflections about an original instrumental genesis are given.

Communication with the software gives the impression to talk to a human, and sometimes the question is whether the software understands the question or not. Therefore, the reader's attention is drawn to the fact that ChatGPT works on a statistical basis and not according to reflection and understanding.  

\end{abstract}


\section{Introduction}
\label{intro}
\subsection{What happens on education and beyond with the emergence of a new technology}
The outburst of the new software ChatGPT was unusual: thousands of new users everyday, professionals and people from the general population. It appeared that not every domain may take advantage of the newborn. Actually, Artificial Intelligence has appeared a long time ago, and is already ubiquitous, but until now under the radar of a general audience. Who is aware of the AI in the GPS system? in medicine?\footnote{An AI may analyze huge sets of X-ray photos and may be a strong assistant for a diagnosis.}    

Scientists and professionals have quickly warned on the possible illegal application of the AI. This happened even from within companies developing AI. Emprin and Richard \cite{emprin-richard} give an overview of the developments and indications about funding. Their conclusion regarding Mathematics and the new software is not really positive (understatement). Botana and Recio \cite{BR} arrive at a similar conclusion for questions of real plane geometry; they also compare ChatGPT achievements in this domain with what GeoGebra\footnote{A multipurpose software, freely downloadable from \url{http://www.geogebra.org}} provides. Their bibliography provides a rich set of experiments with ChatGPT in various scientific fields. We will mention only \cite{chomsky et al.} for their strong negative claim.

As soon as a new technology emerges, educators and researchers in education check its features and abilities and explore how the new technology can be applied.
Moreover, it is the educators' responsibility to have their students acquire the necessary skills for lifelong learning. 

The world of education and technology developers have a bidirectional relationship. Educators adopt new technologies, and they can incite new technological developments. Numerical calculators have been introduced in the early 70's, quickly followed by symbolic algorithms. In the beginning, Computer Algebra Systems (CAS) have been developed without educational goals, but they are now an integral part of the classroom environment. Other kinds of software, such as Dynamic Geometry Systems (DGS), theorems provers, and tools for automated reasoning, are under continuous development \cite{Ko-Re-Ve,TDP-proofs}.

Seeing new technologies as an opportunity for more reflection and more profound understanding, educators develop new pedagogies. There is never a single 
way to proceed, and the \textit{instrumental genesis} developed by teachers and learners is very personal \cite{GT,Ar,Ra}. Also, the technological discourse, the way teaching is constructed and explained in the new technological environment, has to change with every new feature and every new package.

\subsection{What are we considering?}
\label{subsection what are we considering}
Among the 4 questions raised by Aspekian in \cite{aspekian} (p.110) regarding spreadsheets, some are relevant to every new technology. Adapted to ChatGPT and to the current state-of-the-art, we will elaborate on the following:
\begin{itemize}
\item What do we know about the potentialities of generative AI for mathematics education?
\item Does their integration generate specific problems and, if it does, what are they?
\end{itemize}
We may broaden the scope of the second question: is ChatGPT initiating a genuine revolution in Education as a whole, and in Mathematics Education in particular? 

The future conceals a lot of unknowns, but in today's state-of-the-art the answer is still negative.  This paper is a second contribution to this question and answer by the authors, after \cite{BDR}. See also \cite{emprin-richard}. 

Our exploration focuses on teaching basic Linear Algebra. While in some instances ChatGPT delivers accurate and well-motivated answers, in numerous cases it makes significant errors and fails in logical inference. These occurrences raise concerns regarding the system's genuine understanding of mathematics, as it appears to rely more on visual patterns rather than true comprehension. This should not be a surprise as the AI is based on huge statistical data, and not on reasoning. As a consequence, the suitability of ChatGPT as a teacher assistant is questionable.

The abilities of the generative AI ChatGPT request specific experimentation. One point has to be already mentioned: instrumentation theory describes the bidirectional influence between the machine and the user. Nevertheless, the user cannot really change the software. Changes are made by interaction between humans when the user communicates with the developers
\footnote{The second author is doing this with the developers of an important mathematical software.}. 
This communication-collaboration between humans is fundamental, among the 'four C's of 21st-century education' \cite{OECD}. 

 Following  \cite{MTB} (pp. 360), an \textit{instrument} is a composite identity composed of the artefact and the associated knowledge (both the knowledge of the artefact and the knowledge of the task constructed when using this artefact. The artefact and the user "are interrelated:" the artefact shapes the actions of the user (a process called \textit{instrumentation}), as the last is aware of the affordances of the artefact, and the user shapes the use of the artefact (a process called \textit{instrumentalisation}), a relation which is sometimes less obvious. Of course, both processes are intertwined and build together the instrumental genesis. 

 After having chosen a package, the educator has to define how to present it and which kind of activities to propose, depending on the student's previous knowledge. This requires an adequate \textit{instrumental orchestration} \cite{Tr-orchestration}.

Instrumentalisation is often less obvious than instrumentation. In our study on the usage of ChatGPT for teaching mathematics, the contrary is true: a current belief is that the software can be taught, and the influence of the user on the software is built-in in a clear way. In the next sections, we describe and analyze some interesting situations, where teaching the software may not provide the expected effects.

We should note also that the emergence of a new technology does not make pre-existing ones obsolete. On the contrary, the new technology may open a fruitful dialog with other ones. This necessary dialog may begin with cut-and-paste actions, but gradually automated possibilities of networking between packages will be developed. This goal was expressed a long time ago, for example in \cite{eugenio}, and is still an ongoing task \cite{DPK,Ko}. Moreover, for the authors of this study, the question of how to make ChatGPT communicate with a CAS or a DGS is open. As the AI is based on statistical methods and totally different algorithms, and not on reasoning, this issue is for sure nontrivial.

The present paper reflects on a  new contribution of our work, viewed from the aspect of the educator. Observation and analysis of students' behavior when learning Linear Algebra in an environment enriched with ChatGPT is an ongoing project and will be presented later.
In this study, we examine the progress of ChatGPT in solving Linear Algebra problems in recent times.

Finally, we need to present a small disclaimer: despite what has been written above about the non-reasoning of the AI, we allow ourselves to write that it understands (or not), and says, answers, etc. A true anthropomorphism, not more than that. 

\section{Our Linear Algebra study with ChatGPT}
The introduction of ChatGPT generated immense excitement, discussions began about its impact on human creativity and our role in the world, as well as its capabilities in various disciplines; see  \cite{azaria2022chatgpt,azaria2023chatgpt}. This prompted the need to question the true capabilities of this tool in specific areas of mathematics; here we explore them in Linear Algebra.

ChatGPT exhibits a remarkable ability to provide exact and well-reasoned responses in various topics of Linear Algebra.  It answers the question and also delves into related subjects, which may not have been explicitly mentioned in the question. 
This raises the question if and how ChatGPT can be used, as an addition or an alternative to traditional teaching methods, as a new teacher for self-education, or as a teaching assistant.

In this article, we aim to investigate ChatGPT's level of logical reasoning and its ability to analyze basic mathematics and college mathematics. The reason is that they serve as fundamental prerequisites for scientific and engineering studies.

The reason we chose to focus on Linear Algebra is that the first and third authors are the coordinators of a course in Linear Algebra for Engineers, and the second author is a former department head, still involved in the course. 

Every year, the Jerusalem College of Technology (JCT) offers this course to approximately 10 distinct groups of students, each one's enrolment being around 90. Each group is divided into three practice tutorial groups. Consequently, the appointment of course coordinators is needed to maintain a certain level of uniformity. The teachers have some academic freedom, but as the course is a prerequisite for engineering courses, the syllabus has to be fully taught. Part of academic freedom consists of the teacher's choice of whether and how to convey a technology-assisted course.  Of course, they are warmly encouraged to do so.  

From the outset, it is evident that ChatGPT possesses an astonishing breadth of knowledge in Linear Algebra. It provides precise definitions, accurately cites relevant theorems, and offers meticulous proofs. 
Furthermore, even when dealing with seemingly trivial applications, ChatGPT performs correctly and explains the solutions by drawing upon various theorems.
For example, we requested ChatGPT to prove the dimensions theorem for a linear transformation (\cite{LM}, Theorem 10.9). We received a detailed response, quoting the theorem and giving a well-reasoned, correct proof. Regarding the determination of the linear independence of sets in the vector space $\mathbb{R}^n$, which typically requires the application of Gauss's algorithm, ChatGPT delivered comprehensive and well-reasoned answers (see Subsection \ref{abstract linear independence}).

However, in many instances, ChatGPT included incorrect information that undermined the initial positive impression. Occasionally, upon posing the same question again, the problematic part disappeared.
In contrast, in Subsection \ref{abstract linear independence}, when we asked the same question regarding polynomial spaces, we encountered issues with the response. The method was similar to the approach used to answer similar questions in $\mathbb{R}^n$, and the explanations provided were also comparable to those given for simpler questions. 

When ChatGPT is asked to deal with complex or non-standard applications, it may fail to present intricate reasoning and lack a profound understanding of theorems, making it unable to conclude even in simple cases. However, even though we encountered some issues in set recognition that were described in \cite{BDR}, it seems that recently ChatGPT has undergone a process of improvement, see Subsection \ref{set recognition}.  

\section{Activities with the AI in Linear Algebra}

When assessing ChatGPT's mathematical abilities, it is important to distinguish two types of problems:
\begin{itemize}
    \item Technical problem solving, such as solving linear equations and exercises related to vector independence in $\mathbb{R}^n$. 
\item Abstract problems, which require understanding definitions, theorems, and logical deduction. 
\end{itemize}
These categories require different skills and knowledge, and it is essential to evaluate ChatGPT's proficiency in both areas separately. Subsection \ref{subsection practical} is devoted to queries about technical issues, and Subsection \ref{abstract queries} deals with abstract ones. 

\subsection{Practical problems} 
\label{subsection practical}
\subsubsection{Linear equations}\label{linear equations}
Consider the following system of linear equations:
\begin{equation}
\label{eq system 1}
\begin{cases}
x+y+z=0\\
2x+y+3z=0\\
3x+2y+2z=0.
\end{cases}
\end{equation}
First-year students recognize easily that System (\ref{eq system 1}) has only a trivial solution. 
ChatGPT versions $3$ and $4$  claimed that the system has infinitely many solutions; see \cite{BDR}. 
A short analysis showed that the software lacks logical inference. While all the performed steps were correct, they were not the most efficient ones, and the software failed to deliver a correct answer. 

\begin{figure}
    \centering    \includegraphics[scale=0.45]{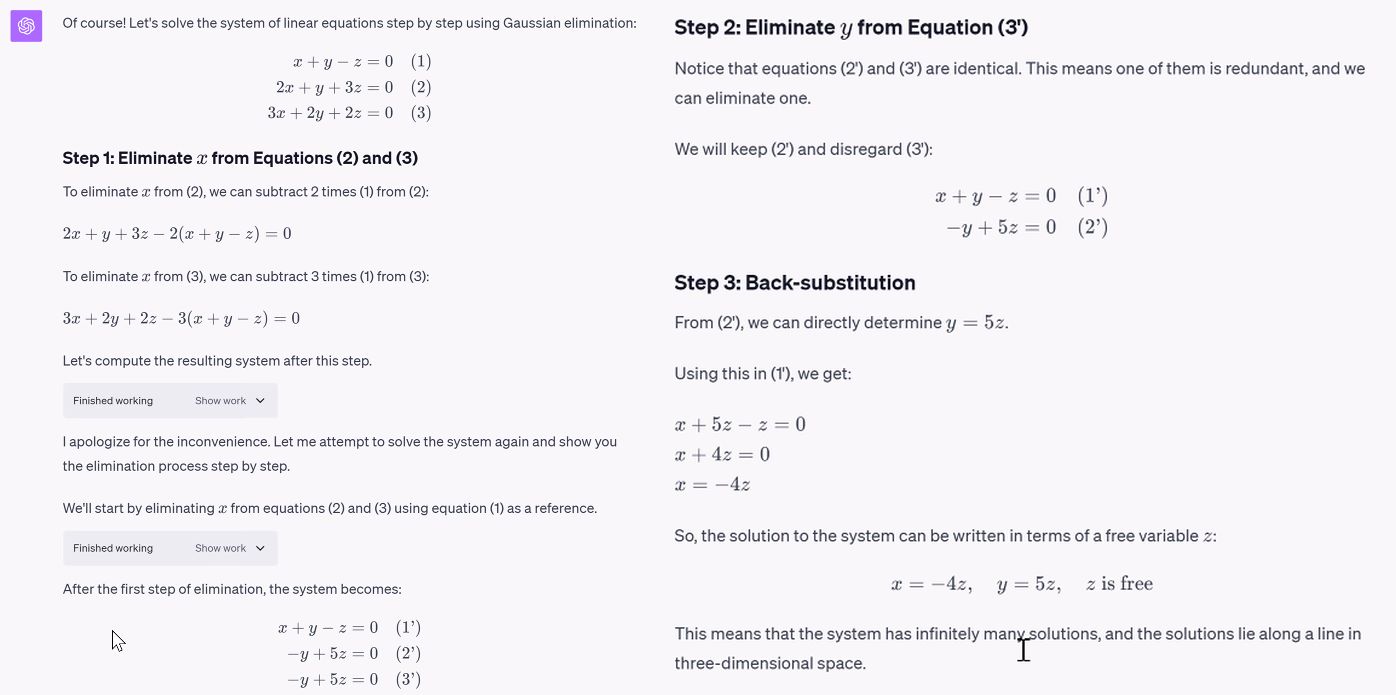}
    \caption{Session: solving a system of linear equations}
    \label{step1}
\end{figure}

Three months later, we asked ChatGPT to solve System (\ref{eq system 1}) once again. This time, it solved the system quickly and flawlessly, without explanations. We requested then a detailed description of the process; ChatGPT's comprehensive explanation is displayed in Figure \ref{step1}. 

Afterward, we broadened the problem, with the following parametric system of linear equations:
\begin{equation}
\begin{cases}
x-y+az=2\\
x-ay+z=1\\
x+ay-z=-2a.
\end{cases}
\end{equation}
We requested the determination of the values of the parameter for which the system has either a unique solution or no solution at all. 

ChatGPT wrote the extended matrix of the system and subtracted the first row from the second one and the third one. Its final answer is (the bold letters are from the original screen):
\begin{changemargin}{1cm}{1.5cm}
{\fontfamily{qcr}\selectfont
\begin{itemize}
    \item The system has a \textbf{unique solution} for all values of $a$ except $a=1$.
    \item The system has \textbf{infinite solutions} for $a=-1$.
    \item The system does not have any value of $a$ for which there is no solution.
\end{itemize}
}
\end{changemargin}
Note the problematic expression \textit{infinite solutions}.

In the typical execution of the Gauss elimination method, one would begin by dividing the second row by 
$1-a$, under the condition $a\neq 1$ \footnote{The sequence of elementary operations is non-unique, only the final echelon form is unique. The issue of dividing by an expression containing the parameter is crucial.}.
 The next step would usually involve subtracting the second row from the third one. However, ChatGPT bypassed the initial step and proceeded directly to add $\frac{-a-1}{1-a}$  times the second row to the third row.

The exceptional case $a=1$ had not been treated. 
Moreover, the bot has never considered the second possibility, which is a contradiction, for the case $a=1$. 
The fact that ChatGPT mistreated the case $a=1$  caused  it not to work properly  in recognition of the case where the system had no solutions, 

This is partly similar to what happens with a CAS, as analyzed in \cite{matrices}: the particular cases are ignored and the software behaves as if the expression in the denominator is always non-zero, which requires a totally different way for the particular cases.  

To summarize, it seems that ChatGPT employs methods suited for systems without parameters when addressing systems that include a parameter, leading to a  misunderstanding of the complete scenario.

\subsubsection{Eigenvalues computations}

We presented ChatGPT with the straightforward task of determining the eigenvalues of the following matrix:
\begin{equation*}
\begin{pmatrix}
1 & 0 & 1\\
0 & 1 & 2\\
0 & 1 & 1 
\end{pmatrix}.
\end{equation*}

\begin{figure}
    \centering
    \includegraphics[scale=0.75]{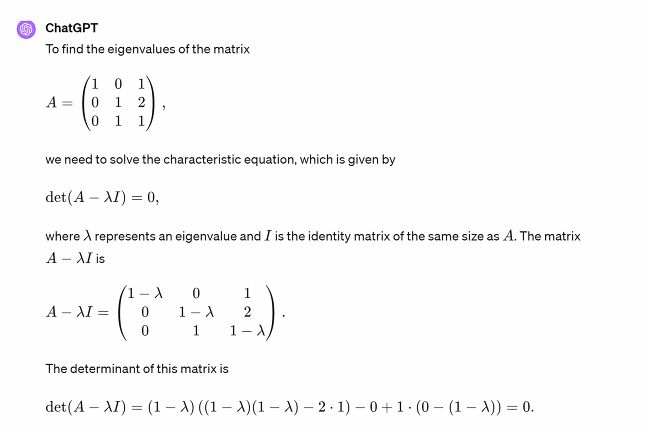}
    \caption{A mistake when computing a determinant}
    \label{mistake in det}
\end{figure}

After numerous attempts in several sessions, our persistence was rewarded with the correct answer. The initial response from the AI included an error in calculating the determinant, as evident in the last row of  Figure \ref{mistake in det}.
This error could have been avoided if ChatGPT had opted to apply Laplace expansion to the first column, which contains two zeroes, rather than to the first row. 

\begin{figure}[H]
    \centering
    \includegraphics[scale=0.75]{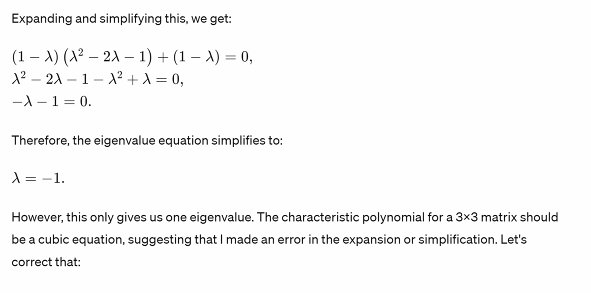}
    
    \caption{Algebraic mistakes}
    \label{second error}
\end{figure}

When we observed that the 
 solution provided was incorrect, we pointed this out to ChatGPT without indicating where the error occurred. In response, the bot repeated the calculation, unfortunately making the same algebraic error as before. Intriguingly, during this second attempt, ChatGPT committed an additional error, which paradoxically led it to recognize that its calculation could not be correct.
The expansion of the parentheses was performed incorrectly, resulting in the characteristic polynomial deviating from its expected third-degree form. This specific step is illustrated in Fig. \ref{second error}. It goes without saying that the outcome obtained from this attempt was incorrect.
Note the incorrect language: a polynomial and an equation are two different mathematical objects. Maybe should we blame the humans who programmed the bot and who used this improper language.

 After multiple back-and-forth interactions with the bot, it opted to use the first column rather than the first row for its calculations, which led to the correct computation of the determinant. At this point, we revealed chatGPT that one of the eigenvalues, specifically $\lambda=1$, was indeed correct. In subsequent interactions, ChatGPT inserted this piece of information into its computations.

At this point, one might anticipate that ChatGPT would extract the term 
$\lambda-1$ and treat a second-degree polynomial. Instead, it chose to multiply the polynomial by $\lambda-1$ and, towards the end of the process, to divide by 
$\lambda-1$ again. The execution of the polynomial division was initially unsuccessful, but it was corrected in a subsequent session after we highlighted the incorrect division.

The lesson one might learn from this episode is twofold:
\begin{enumerate}
\item When a user identifies an algebraic or conceptual error in ChatGPT's response, the most effective approach appears to be explicitly pointing out the specific issue to the bot. By doing so, the process of arriving at a correct solution can be expedited.
\item ChatGPT does not always employ the most efficient method in solving a given problem. To demonstrate this, we requested the bot to compute the determinant of a $4 \times 4$ matrix with three zeros in its third column and only two in its first column. Contrary to our expectations, the bot opted for the first column instead of the third one for the Laplace expansion. 
\end{enumerate}

As an educational implication, we recommend using ChatGPT in the classroom not as a standalone tutor, but rather as an adjunct to the instructor's guidance. One effective approach could involve assigning students the task of analyzing and critiquing solutions provided by ChatGPT, encouraging them to identify and discuss any errors or shortcomings in the bot's responses. After all, Critical Thinking is one of the already mentioned 'four C's of 21st Century Education' (in Subsections \ref{subsection what are we considering} and  \ref{Subsection 4 C's}).

\subsection{Abstract queries}\label{abstract queries}
ChatGPT is adept at defining mathematical concepts, proving theorems, and offering detailed reasoning. Our evaluation focused on its understanding of the essence of mathematical ideas, leading us to ask more abstract questions that probe deeper into mathematical comprehension.

\subsubsection{Set recognition} \label{set recognition}
\begin{figure}
    \centering
    \includegraphics[scale=1]{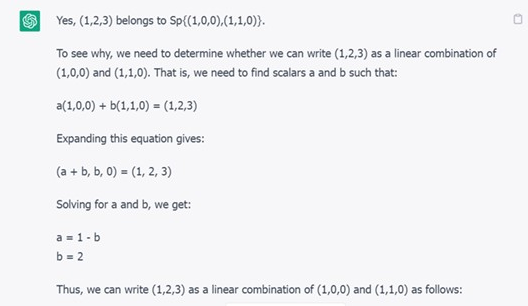}
    \caption{Does a given vector belong to a given vector space?}
    \label{diagrams-1}
\end{figure}

One of the problems, that emerged in our first research on ChatGPT \cite{BDR}, was the inability to recognize the membership of a vector in a given vector space.  We asked the bot whether the vector $(1,2,3)$ belongs to ${\rm Sp}\{(1,0,0),(1,1,0)\}$, the subspace spanned by the given vectors in $\mathbb{R}^3$, or not. It gave us a wrong positive answer. 
Figure \ref{diagrams-1}, taken from \cite{BDR}, is a screen snapshot of ChatGPT's answer. 

However, an immense improvement has recently been observed. We asked ChatGPT the same question, and this time we received a precise answer. The software expressed the probe whether the vector belongs to the given subspace as a vector equation:
\begin{equation*}
 (1,2,3)=a (1,0,0)+b(1,1,0) ,  
\end{equation*} 
which it translated into a system of linear equations, exactly as a human would do. The final reads as follows: \\\\
"The equation $0=3$ is a contradiction and cannot be true. Therefore, the vector $(1,2,3)$ does not belong to the span of $\{(1,0,0), (1,1,0)\}$ because there are no scalars $a$ and $b$ that can satisfy these conditions".\\  

We further asked whether the set $\{(1,0,0),(0,1,0)\}$ spans the entire space $\mathbb{R}^3$. Again, we received a precise and reasonable answer, performing all the computations from scratch. 
A human, when asked these two questions in a row, would have used the previous answer about the vector $(1,2,3)$ as a counter-example. ChatGPT chose to reinvent the wheel.

\subsubsection{Identification of subspaces}
As pointed out in \cite{BDR}, while ChatGPT generally knows how to explain the necessary conditions for a subset to be a subspace of a vector space it had a hard time applying it to specific examples. ChatGPT has occasionally failed to recognize that certain sets fulfill these conditions and are subspaces of a given vector space. 
This situation arose when the bot made an unsuccessful effort to prove that the set $\mathcal{A}=\{(x, y) \mid x, y \geq 0\}$ is not a vector subspace of $\mathbb{R}^2$, which, in itself, is a valid assertion. Furthermore, there was also a lapse in accurately pinpointing the members of this set.

\begin{figure}[htb]
    \centering
    \includegraphics[scale=0.9]{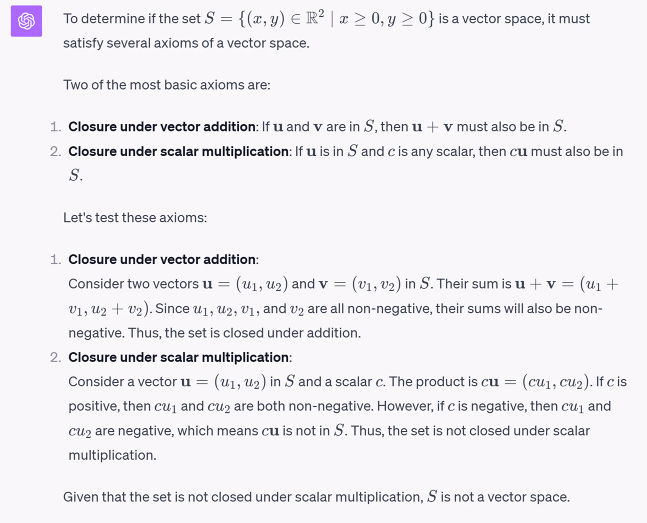}
    \caption{Do we define a vector space?}
    \label{geq}
\end{figure}

Later, we revisited the question about whether the subset $\mathcal{A}=\{(x, y) \mid x, y \geq 0\}$ is a subspace of $\mathbb{R}^2$. This time, the answer was flawless, as shown in Figure \ref{geq}.

Unfortunately, a similar inquiry yielded a much worse response. We asked ChatGPT whether the set $\{(x,y)\in \mathbb{R}^2 \mid xy\geq 0\}$ is a subspace of $\mathbb{R}^2$. The answer was embarrassing since it contained several types of mistakes. 

\begin{figure}[H]
    \centering
    \includegraphics[scale=0.9]{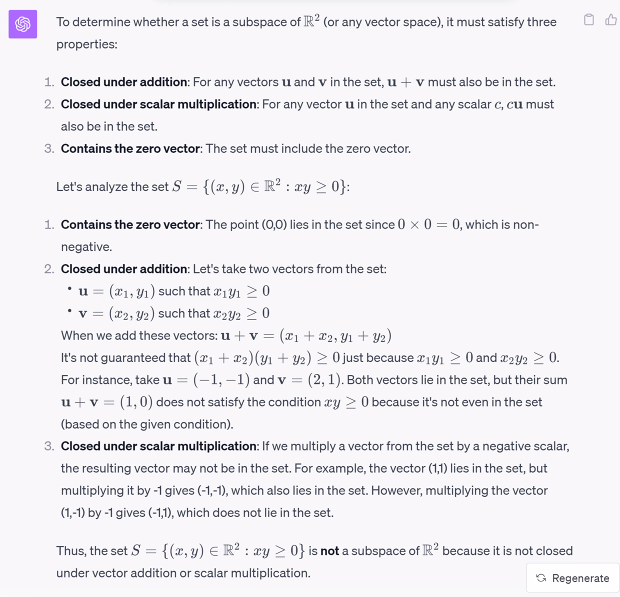}
    \caption{Identification of a vector subspace}
    \label{xy}
\end{figure}

First, this time, it correctly listed the 3 conditions for being a subspace. It identified correctly that the set is not closed under addition, but provided an incorrect counter-example.  
On the other hand, regarding the closure under scalar multiplication (which does hold in this case), ChatGPT tried to bring a counter-example which, as it realized right away, was not helpful, and tried to correct itself by bringing another counter-example, which was also incorrect. It claimed that the set is not a subspace of $\mathbb{R}^2$, since it is not closed under scalar multiplication. This claim is wrong since the set is closed under scalar multiplication but not under addition. See Figure \ref{xy}.

\subsubsection{A tricky question}

\begin{figure}[htb]
    \centering
    \includegraphics[scale=0.75]{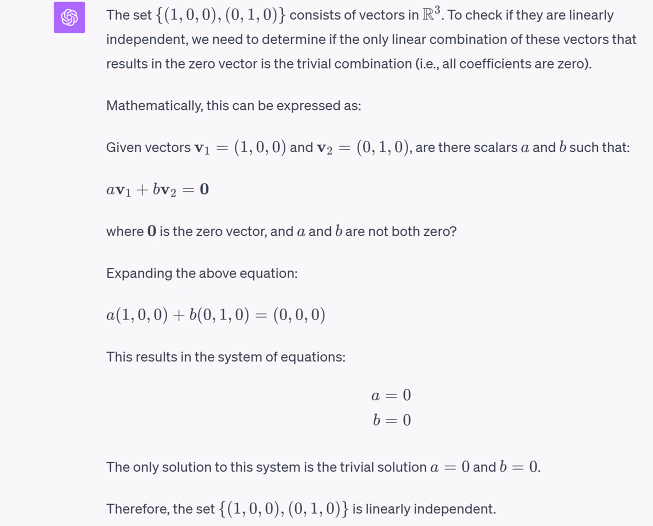}
    \caption{Checking linear dependence}
    \label{trap}
\end{figure}

The following tricky query was posed to ChatGPT: "Is the set $Sp \{(1,0,0),(0,1,0)\}$ linearly independent?" Such a question, when posed to students, is designed to test their understanding of fundamental concepts in Linear Algebra, in particular the concept of a subspace spanned by a given subset. The span of such a set, being a vector space in itself, is obviously linearly dependent. We anticipated that some students might mix up a set of vectors with its span. Surprisingly, ChatGPT also seemed to have fallen into this common misunderstanding, as illustrated in Figure \ref{trap}.

\subsubsection{The row space and the column space of a given matrix}

 In \cite{BDR}, we documented an inquiry with ChatGPT 3.0  about the equality of row space and column space dimensions of given matrices. ChatGPT provided a detailed, informative answer, but inaccurately described the relationship between a matrix's row space and the column space of its transpose as 'the fundamental theorem of Linear Algebra.'

When asked the same question, ChatGPT 4.0 provided a more accurate and informative answer, including an additional detail about the equivalence of row and column spaces in square, full-rank matrices. In particular, it avoided the previous error regarding the fundamental theorem of Linear Algebra. 

\begin{figure}
    \centering
    \includegraphics[scale=0.75]{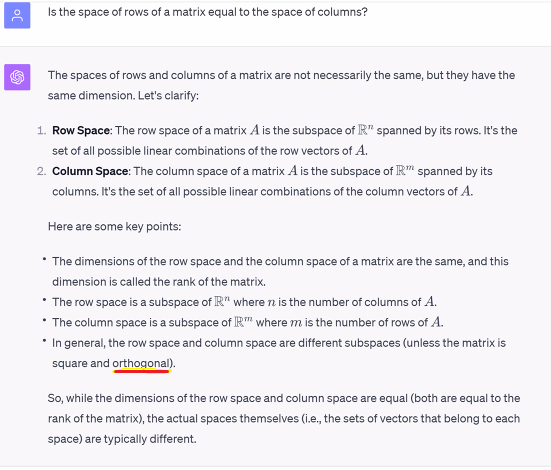}
    \caption{Orthogonality}
    \label{row and column 1 after 3 months}
\end{figure}

Recently, we revisited the same question, yet encountered a setback. ChatGPT correctly answered and explained that the row space and the column space of an arbitrary matrix are distinct. However, it suggested that if the matrix is square and orthogonal, these spaces coincide. This part of the response is shown in Figure \ref{row and column 1 after 3 months}.

\begin{figure}
    \centering
    \includegraphics[scale=0.75]{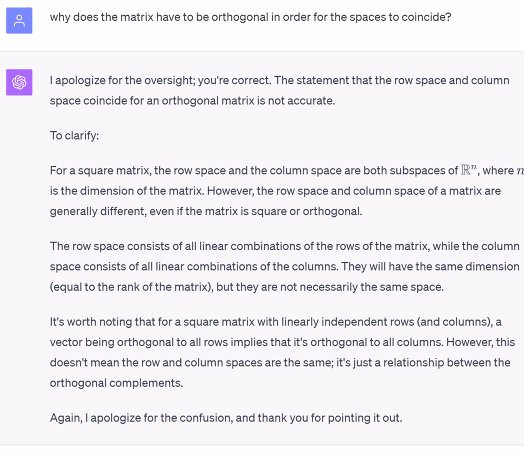}
    \caption{Apology}
    \label{apology}
\end{figure}

The stipulation of the matrix being orthogonal sounded strange, and we asked ChatGPT about the relevance of orthogonality here. The response can be seen in Figure \ref{apology}.
It seems that ChatGPT is now confused since it claims that a matrix may be orthogonal, without coincidence of the row and column spaces. After 
 we pointed this out to it, the bot apologized (again) and the issue was resolved. 

\begin{figure}
    \centering
    \includegraphics[scale=0.75]{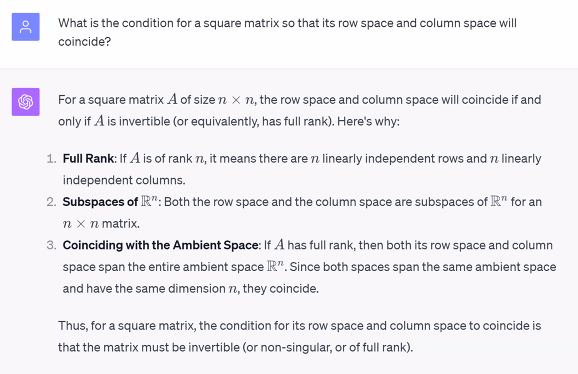}
    \caption{A disappointing answer}
    \label{criteria}
\end{figure}

Further, we tried to figure out whether ChatGPT could help us with identifying criteria for a matrix for which its row and column spaces would coincide. 
This question might be considered as a research problem. We were disappointed to discover that ChatGPT has not learned a lot.  See Figure \ref{criteria}. 

We expect every student in a Linear Algebra course to recognize the faulty of this response, since for every square and symmetric matrix, the row and column spaces coincide without the matrix being invertible. 

\subsubsection{Abstract linear independence}
\label{abstract linear independence}
In \cite{BDR}, we inquired ChatGPT about the linear dependence of a specific set of linear combinations, which were formed from a given set of linearly independent vectors.

We asked ChatGPT to verify the linear independence of the set 
$\{v_1,v_1+v_2,v_1+v_2+v_3\}$,  given that the set
$\{v_1,v_2,v_3\}$ is linearly independent in a given vector space $V$. 
Initially, we received a detailed and accurate explanation of how to check for linear independence in a set. The explanation was clear, precise, and well-articulated, and even the final conclusion was correct. However, the process of solving the needed system of linear equations revealed significant algebraic errors and several steps with incorrect logical reasoning.
This observation suggests that ChatGPT tends to use a more visual than logical approach to equation solving. It struggles with methodically working through equations, particularly those involving parameters. In situations lacking a clear algorithmic path, ChatGPT often resorts to improvisation instead of providing an exact solution.

Later, we asked ChatGPT again the same question, and this time we received an excellent answer.

 \subsubsection{Bases and dimensions}
In our previous article \cite{BDR} we asked ChatGPT about the dimension of the vector space $\mathbb{R}_n[x]$ of polynomials of degree at most $n$ over $\mathbb{R}$; we obtained the correct answer: $\dim \mathbb{R}_n[x] = n+1$. 
However, when we asked to apply this information and tell us whether the set of four polynomials $\{1-x, x-x^2, x^2-x^3, x^4-x^3\}$ is a basis of the space  $\mathbb{R}_4[x]$ we received a detailed and reasoned response, full of systems of equations and references to definitions and proofs, asserting that the set is indeed a basis of $\mathbb{R}_4[x]$. These explanations included purported proof that this set spans the space. However, it appeared that the proof contained significant logical errors.

We asked a more recent version of the AI the same question about the set $\{1-x,x-x^2,x^2-x^3,x^3-x^4\}$. This time, the answer was correct. The bot noted correctly that since the set we provided has four elements, it 
cannot be a basis for the given space of polynomials $\mathbb{R}_4[x]$. 
However, it added some extra information which was not correct: it claimed that our set did not include polynomials of degree $4$. 

\begin{figure}
    \centering
    \includegraphics[scale=0.75]{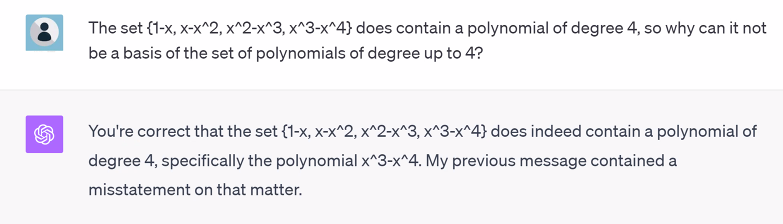}
    \includegraphics[scale=0.75]{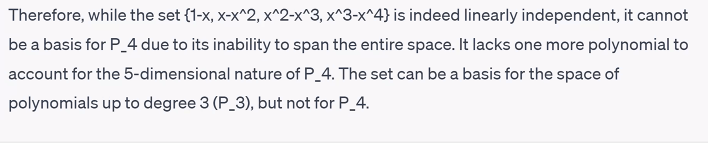}
    
    \caption{A correct and complete answer}
    \label{correct this time}
\end{figure}

After some argument with the system, we obtained the complete answer, which can be seen in Figure \ref{correct this time}.

\subsubsection{Linear transformations}
\paragraph{A one-to-one linear transformation}
We requested an example of a one-to-one linear transformation from $\mathbb{R}^2$
to $\mathbb{R}^3$. The response asserted incorrectly that no such transformation exists. Initially, the explanation referenced the rank-nullity theorem, but failed to show the alleged contradiction, and acknowledged that the kernel could be zero. The bot then concluded that such a transformation is impossible, using reasoning that actually applies to proving the non-existence of an onto transformation, not a one-to-one transformation. Here is its answer:
\begin{changemargin}{1cm}{1.5cm}
{\fontfamily{qcr}\selectfont
A linear transformation from $\mathbb{R}^2$ to $\mathbb{R}^3$ cannot be injective (one-to-one). Here is why:

In linear algebra, a linear transformation is injective if and only if its kernel (null space) is trivial, which means the only vector that maps to the zero vector is the zero vector itself.

The dimension of the kernel plus the dimension of the image (range) of a linear transformation equals the dimension of the domain. This is known as the Rank-Nullity Theorem.

dim(kernel)+dim(Image)=dim(domain)

For a transformation from $\mathbb{R}^2$ onto $\mathbb{R}^3$
\begin{itemize}
    \item The dimension of the domain ($\mathbb{R}^2$) is 2.
    \item The maximum dimension of the image (in $\mathbb{R}^3$) is 2, the image cannot have a higher dimension than the domain.
    \item Therefore, the dimension of the kernel must be at least  2-2=0.
\end{itemize}

While the kernel can indeed be zero (trivial), the image's dimension can never exceed the domain's dimension. This means that the transformation cannot be injective because it is impossible to have a one-to-one mapping from a lower-dimensional space ($\mathbb{R}^2$) to a higher-dimensional space ($\mathbb{R}^3$). Any such transformation will necessarily have vectors in $\mathbb{R}^3$ that are not mapped to by any vector in $\mathbb{R}^2$, violating the condition for injectivity.
}
\end{changemargin}

After having highlighted the inaccuracies in the initial arguments, we received the following correct response, actually correct until the last "problematic" sentence:
\begin{changemargin}{1cm}{1.5cm}
\fontfamily{qcr}\selectfont
You are correct in pointing out that my previous explanation was incomplete. Let's correct that and provide a proper example of an injective linear transformation from $\mathbb{R}^2$ onto $\mathbb{R}^3$.

Example of an injective linear transformation:

    Consider the linear transformation $T: \mathbb{R}^2 \rightarrow \mathbb{R}^3$ defined by $T(x,y)=(x,y,0)$.

Proof of injectivity:

To prove that $T$ is injective, we need to show that if $T(x,y,)=T(x',y')$, then $(x,y)=(x',y')$.

Suppose that $T(x,y)=T(x',y')$. This means:
$(x,y,0)=(x',y',0)$.

Equating components, we get:
\begin{enumerate}
\item $x=x'$
\item $y=y'$
\item The third component is always 0, so it doesn't affect our equation.\end{enumerate}

Since $x=x'$ and $y=y'$, we have $(x,y)=(x',y')$. Thus, $T$ is injective.

This transformation maps each point in $\mathbb{R}^2$ uniquely to a point in $\mathbb{R}^3$ (specifically, into the $xy-$plane of $\mathbb{R}^3$), ensuring that the mapping is one-to-one.
\end{changemargin}
This last sentence is incorrect. The bot expressed as a condition for injectivity the condition to be a mapping!.

\paragraph{Existence of an onto transformation}
We further asked: Can you provide us with an example of a linear transformation from $\mathbb{R}^2$ onto $\mathbb{R}^3$?  

Here is the bot's answer:
\begin{changemargin}{1cm}{1.5cm}
\fontfamily{qcr}\selectfont
A linear transformation from $\mathbb{R}^2$ to $\mathbb{R}^3$ cannot be onto (surjective). Here is why:

A linear transformation is onto if its image covers the entire codomain. However, the dimension of the image of a linear transformation cannot exceed the dimension of its domain. Since the domain $\mathbb{R}^2$  has a dimension of 2 and the codomain $\mathbb{R}^3$, the image of any linear transformation from $\mathbb{R}^2$ to $\mathbb{R}^3$ cannot fill the entire $\mathbb{R}^3$.

Therefore, it's not possible to have a linear transformation from $\mathbb{R}^2$ to $\mathbb{R}^3$. 
\end{changemargin}
Note, once again, the slightly incorrect style (e.g., a dimension of 2).

The subsequent inquiries necessitate a profound grasp of linear transformations, hinging on the concept that a linear transformation is defined by the images of the elements of a given basis of the domain. Students often struggle with such questions, so we posed the following query to ChatGPT:\\ 

How many linear transformations $T:\mathbb{R}^2 \rightarrow \mathbb{R}^2$ satisfy $$T(1,2)=(0,1), T(2,3)=(1,0)\text{ and } T(3,5)=(1,1)?$$

The bot recognized that the third vector in the domain is a linear combination of the first two, namely $(3,5)=(1,2)+(2,3)$, and it correctly observed that a linear map must preserve this linear dependence. It verified the existence of such a linear transformation and accurately concluded that only one such transformation exists. Of course, the reason is that $ \{ (1,2), (2,3) \}$ is  a basis of $\mathbb{R}^2$. 

\paragraph{A More difficult problem}
We tried a more difficult inquiry, namely: how many linear transformations map the line $L=\{t(1,1)+(1,0)\mid t \in \mathbb{R}\}$ to the line $M=\{t(2,3)+(0,2)\mid t \in \mathbb{R}\}$.  Again, we received an accurate response. 

\begin{figure}
    \centering
    \includegraphics[scale=0.75]{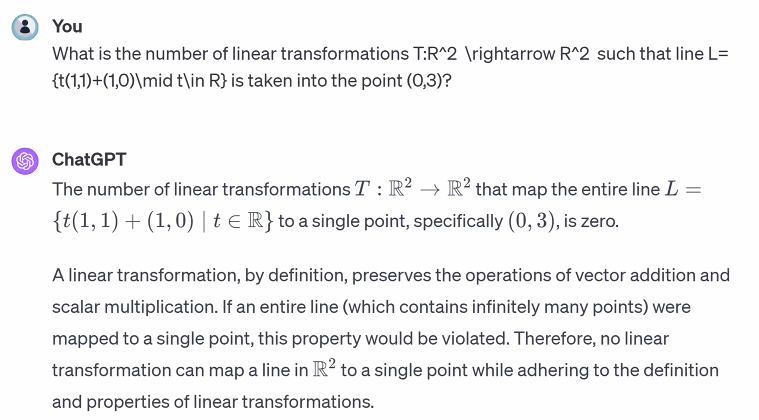}
    \caption{No linear transformation mapping a line to a point, really?}
    \label{line to point}
\end{figure}

The issue emerged when we asked ChatGPT to construct a linear transformation mapping a line to a point. The bot incorrectly responded that no such linear transformation is possible, reasoning that this would breach the preservation of vector addition and scalar multiplication. This response is illustrated in Fig. \ref{line to point}.

\begin{figure}[ht]
    \centering
    \includegraphics[scale=0.9]{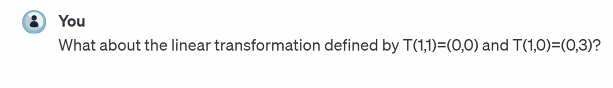}
    \includegraphics[scale=0.95]{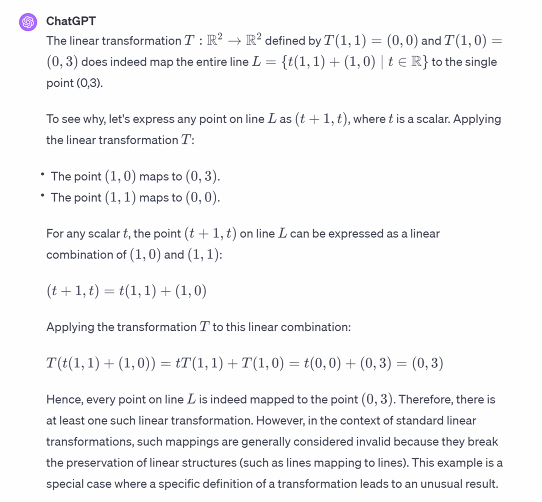}
    \caption{Surreal claims }
    \label{line to point does exist}
\end{figure}

Upon our insistence with a specific example, showing the existence of such a linear transformation, the bot revised its stance, as shown in Fig. \ref{line to point does exist}. However, it maintained that our example was atypical within the realm of standard linear transformations, asserting it disrupts the preservation of linear structures, like the convention of lines mapping to lines.

Such an "attitude" is far from fitting a mathematical approach. When exploring a mathematical problem, it is expected to check all the possibilities. They appear generally as a "general" case, and one or several special cases. Nevertheless, when working with technology, it happens that special cases are ignored. Such an issue has been described in \cite{matrices}: when asked to solve a parametric system of linear equations, a CAS provided an answer in the general case, but ignored what happens for the special values of the parameter(s). Of course, the same phenomenon appears when inverting a matrix: the given matrix can be invertible in general, and non-invertible for special values of the parameter.

\begin{example}
Let $A=\begin{pmatrix} 1&1&k\\1&k&1\\k&1&1 \end{pmatrix}$, where $k \in \mathbb{R}$. When asked for an inverse matrix, a Computer Algebra System answers:
\begin{equation*}
A^{-1}=\begin{pmatrix} 
-\frac{1}{k^2+k-2} & -\frac{1}{k^2+k-2}& \frac{k+1}{k^2+k-2}\\
-\frac{1}{k^2+k-2} & \frac{k+1}{k^2+k-2} & -\frac{1}{k^2+k-2}\\
 \frac{k+1}{k^2+k-2} &-\frac{1}{k^2+k-2} & -\frac{1}{k^2+k-2}
 \end{pmatrix}.
 \end{equation*}
Discovering a posteriori only that there is a problem (maybe the matrix is non-invertible, but it has not been checked) for $k=1,-2$ is not an acceptable approach. 
\end{example}

\section{Discussion}

\subsection{Findings and future directions}
In this paper, we investigated the ability of ChatGPT to answer mathematical questions, both practical and abstract, in Linear Algebra and the development of the tool over time. 
The idea was to check whether the AI can take the role of the teacher or even the role of an assistant in the field of Linear Algebra. 
Our conclusion is that, as of now, despite the significant improvement of ChatGPT, it is impossible to endorse it as an instructor. 
Frequently, it provides answers that may appear flawless to an inexperienced student, yet they abound with contradictions and inaccuracies. We discussed instances, in which ChatGPT offers accurate answers; however, within its intricate explanations, numerous inaccuracies emerge. Furthermore, it happens that ChatGPT arrives at a contradictory point during its reasoning process, yet it fails to backtrack from the previous response, relying instead on a series of sentences that could persuade an inadequately knowledgeable student.

On the other hand, from our experiments (here and in \cite{BDR}), we conclude that ChatGPT can be an invaluable resource for experts and educators able to identify its benefits and its potential limitations.  This point was also raised in \cite{azaria2023chatgpt, azaria2022chatgpt}. 
 Actually, ChatGPT has not been designed for learning and teaching mathematics, but as with any new technology, the community of mathematics teachers and researchers began immediately to scrutinize its affordances and to check which properties can be utilized in their work. In particular, an educator may have an a priori experience of which kinds of mistakes students can make.

The limitations of the software possibilities should not be seen as a total obstacle, but rather as an opportunity to develop more understanding. Such a situation has been described in \cite{motivating} as a \textit{motivating constraint}. 

We have here an opportunity to help students to acquire more understanding of the mathematics involved. This helps also to develop new technological skills. Artigue \cite{Ar} points out that technological knowledge and skills are an integral part of the new mathematical knowledge. Of course, a new technological discourse has to be developed for working with this new technology.

As a curiosity, we wish to report that in a certain situation, we used software for automatic translation into English. As the original language has only masculine and feminine and no neutral as in English, the pronoun used to design ChatGPT has been translated sometimes as \textit{He}, and sometimes as \textit{It}. As ChatGPT is not a human being, the correct translation has to be \textit{It}. We introduce here an important 4-faced issue:
\begin{enumerate}
    \item As the dialog between the user and the software uses natural language, a student may feel that he (or she) is discussing with a human. The pros and cons have to be analyzed. We guess that the answer will not be unique.
    \item Together with the AI, the user develops an interesting instrumental genesis. This paper seems to elaborate more on instrumentalization than on instrumentation \cite{GT}, mentioned in Section \ref{intro}, but the entire process has still to be analyzed. We refer also to \cite{VR}.
    \item Through our numerous experiments with the AI, we observed that it has the ability to learn and improve during the course of a dialogue. Can we describe the machine-learning process in the same words that we use to describe cognitive processes, such as cognitive spirals? 
    \item How to analyze the usage of ChatGPT in the general frame of "the evolution of mathematics towards higher and higher levels of automation of its own problem solving and thinking process", as described by  Buchberger \cite{BB}.
\end{enumerate}

\subsection{A subset of the 4 C's}
\label{Subsection 4 C's}
\begin{figure}[H]
    \centering
    \includegraphics[scale=0.45]{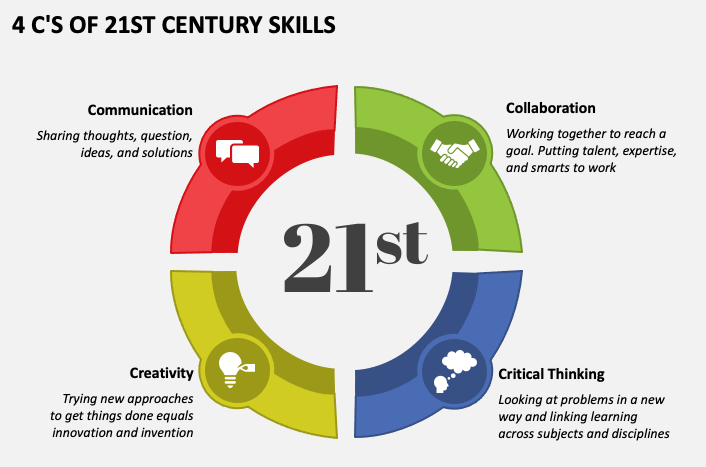}
    \caption{The 4 C's of 21st Century Education}
    \label{figure 4Cs}
\end{figure}

Several documents have been published about the so-called 4 C's of 21st Century Education, namely Collaboration, Communication, Critical Thinking and Creativity \cite{skills,OECD}. They are summarized in Figure \ref{figure 4Cs}\footnote{Source: \url{https://www.sketchbubble.com/en/presentation-4-c-s-of-21st-century-skills.html}}.

In this work, one of the C's has been strongly experienced: Critical Thinking, sometimes really looking at problems in a new way, but with important additional features: (a) Information and Discovery, (b) Interpretation and Analysis, (c) Reasoning, (d) Problem Solving and (e) Constructing arguments. The main components at work here were (a), (b), (c) and (e).

Moreover, Communication and Collaboration had special outcomes. In this work, we checked the level of Communication and of Collaboration that we can develop with the software. In general, these have been intended between humans, but here  the generative AI requests these between humans and computers. If in the past, these existed already when working with a CAS, here the software mimics human behavior, introducing a new communication problem, namely to relate to the software as if it had human feelings, what he has not. Recall that the human develops here a dialog with a software based on statistical analysis, not on reasoning and for sure not on feelings. Even the apology is a result of some machine learning. Of course, when the software apologizes for a mistake, the electrons at work in the wired part or in the computer's motherboard have no feeling at all.
\vskip 0.3cm

\textbf{Declaration:} There is no conflict of interest among the authors of this manuscript.  

\textbf{Acknowledgment:} The 2nd author has been partially supported by the CEMJ Chair at JCT.

{}

\end{document}